\documentclass[runningheads]{llncs}
\usepackage{graphicx}
\usepackage{subfig}
\usepackage{caption}
\usepackage{amstext}

\begin{document}

\title{Improving Primate Sounds Classification using Binary Presorting for Deep Learning}

\titlerunning{Primate Sounds Classification using Binary Presorting}

\author{
Michael Kölle\inst{1} \and
Steffen Illium\inst{1} \and
Maximilian Zorn\inst{1} \and
Jonas Nüßlein\inst{1} \and
Patrick Suchostawski\and
Claudia Linnhoff-Popien\inst{1}
}

\authorrunning{M. Kölle et al.}

\institute{Institute of Informatics, LMU Munich, Oettingenstraße 67, Munich, Germany
\email{\{michael.koelle, steffen.illium, jonas.nuesslein, linnhoff\}@ifi.lmu.de}}

\maketitle

\begin{abstract}
In the field of wildlife observation and conservation, approaches involving machine learning on audio recordings are becoming increasingly popular.
Unfortunately, available datasets from this field of research are often not optimal learning material; Samples can be weakly labeled, of different lengths or come with a poor signal-to-noise ratio.
In this work, we introduce a generalized approach that first relabels subsegments of MEL spectrogram representations, to achieve higher performances on the actual multi-class classification tasks. For both the binary pre-sorting and the classification, we make use of convolutional neural networks (CNN) and various data-augmentation techniques.
We showcase the results of this approach on the challenging \textit{ComparE 2021} dataset, with the task of classifying between different primate species sounds, and report significantly higher Accuracy and UAR scores in contrast to comparatively equipped model baselines.
\keywords{audio classification, binary presorting, relabelling, background noise, CNN}
\end{abstract}

\section{\uppercase{Introduction}}\label{sec:introduction}
Along the worldwide loss of biodiversity, measures to preserve wildlife are becoming increasingly more important. Life on earth is largely determined by its diversity, but also by the complex interaction of fauna and flora (cf. \cite{schulze2012biodiversity}). However, to take the necessary protective measures, thorough research is needed to gain knowledge about our environment.
One crucial part of this process often involves long-term observation, e.g., of the behavior and development of a wide variety of wild animals. By continuous documentation, possibly harmful changes can be detected at an early stage and the necessary protective countermeasures can be taken.

Among other data samples like movement patterns, population densities, social structures, the collection of acoustic data is often used to detect, recognize, locate and track animals of interest \cite{blumstein2011acoustic}. Wildlife observation through audio recordings bear several advantages, in contrast to, e.g., camera traps~\cite{swann2014camera}:
    \textbf{1.} they do not have to be precisely aimed or accurately placed and do not need to be moved as often;
    \textbf{2.} group behavior of animals can be studied, even in inaccessible habitats, like underwater or in dark spaces, while introducing no intrusion; 
    \textbf{3.} fewer amounts of data are generated, which reduces complications in storing, securing, distributing, or processing the data \cite{sundaresan2011management,harris2010automatic,hamel2013towards}.

Recent advanced in machine learning~(ML) fueled the involvement of evaluating such wildlife audio recordings. However, for ML models to learn efficiently and consistently, clean data of adequate size is important.
Audio-based datasets, specifically those recording high-noise environments (primate habitats) in the jungle, are made all the more challenging due to misleading noises (e.g., birds that mimic their acoustic surrounding), a poor signal-to-noise ratio (SNR, S/N) from strong background noise or electronically introduced inferences \cite{heinicke2015assessing}. General problems like irregular class distribution (underrepresentation), varying sample length or `weak labeling‘ further increase the analysis difficulty.
While it is still possible to reduce noise in recordings and classify specimen correctly \cite{dogra2021noise}, for such methods to work robust, sufficient pre-labeled training datasets are needed.
The processes to generate these labels (human hearing, crowed-sourcing) can unfortunately be error-prone on their own, which often results in partially mislabeled datasets~\cite{lin2014re}.

In this work, we address these data problems and present an approach which aims to achieve more accurate ml-model predictions for noisy and weakly labeled audio datasets.
Our approach is then evaluated on the `Central African Primate Vocalization dataset'~\cite{zwerts2021introducing} (as part of `ComparE 2021' Challenge~\cite{schuller2021interspeech}).
The goal here is, to develop a multi-class classification model that can distinguish between the sounds of different primates specimen and pure jungle noise (`background').
For this purpose, we combine several methods, such as data-augmentation, pre-sorting by binary pre-sorting and thresholding on the scope of smaller segments of the original files.

The work at hand is structured as follows: We start by describing the dataset, it's challenging aspects and the concepts behind them in Section~\ref{sec:primate_dataset}. Details on the methods we use are specified in Section~\ref{sec:methods}. Our results are presented in Section~\ref{sec:experiments}, which we then differentiate to comparable and related work in Section~\ref{sec:related_work}, before concluding in Section~\ref{sec:conclusion}.

\section{\uppercase{Related Work}}
\label{sec:related_work}
Before the introduction of the used methods, we utilize this chapter to briefly compare ours – Binary pre-sorting and the CNN architecture for audio classification – to established work in the literature.

\subsection{Pre-Sorting}
Lin et al., were first to show the usefulness of pre-sorting when the underlying dataset has a high number of features and, consequently, distributional inconsistencies in the samples \cite{lin2014re}. We therefore consider the challenging ComparE data a natural fit for this technique.
The idea of adjusting data labels (pre-sorting) during the training process can also be found in the work of \cite{wu2018light}.
For this purpose, they use a combination of different methods, consisting of a modified, generalized Rectified Linear Unit (ReLU) activation function -- called MaxOut -- in the CNN, three different networks for training, and (later) `semantic bootstrapping', cf. \cite{wu2016semantic}. Semantic bootstrapping describes a procedure in which certain patterns are extracted from already learned information to gain more and more detailed information. 
Semantic bootstrapping on the data features is performed iteratively on increasingly more general patterns with a pre-trained deep network, so that later during the pre-sorting inconsistent labels can be detected and subsequently changed (or individual, unhelpful samples removed completely). 
In our work, we also employ a pre-trained model for the pre-sorting, but we substitute the notion of semantic bootstrapping with the binary classification task as our preferred way to learn a generalized feature understanding.
A similar approach to this is also taken by \cite{dinkel2022pseudo}. Here, a CNN classifier was trained on a weakly labeled audio dataset. Shorter audio segments were used for training and pseudo-strong labels were assigned to them. Then, the probabilities of the predictions are compared with the original and pseudo-labels to determine the new label. This corresponds to the approach in this work, i.e., the majority-vote of multiple binary-classifiers. For the general multi-class training, Dinkel et al. did not use any form of augmentation, however.
\cite{iqbal2020learning} were also able to show that pre-sorting data can have a significant positive effect on training. They trained an auxiliary classifier on a small data set and used it to adjust labels. It was shown that only a small amount of data is needed to train a meaningful auxiliary classifier. However, unlike the dataset in this work, their samples have labels with a ground truth, so the auxiliary classifier could learn much more confidently than in our case.

\subsection{Audio Classification with CNNs}

To extract relevant features from audio files, it is possible to use the raw audio signal~\cite{lee2017sample,zhu2016learning} or to extract 2D spectrograms in image form from audio files~\cite{wang2019music,nasrullah2019music,choi2016automatic} to generate audio features from the image representation.

The application of CNNs for audio ML tasks has consistently shown promising results, from audio classification, e.g., \cite{hershey2017cnn,muller2020acousticanomaly,muller2020acousticleak}, speech recognition or audio recognition in general \cite{palanisamy2020rethinking,huang2018aclnet}. Contrasting the pure CNN approach is work utilizing recurrent neural networks (RNNs)~\cite{dai2016long,muller2021deep,gimeno2020multiclass}, and, more recently, a combination of both; Convolutional Recurrent Neural Networks (CRNNs)~\cite{wang2019music,adavanne2017sound,choi2017convolutional,cakir2017convolutional} also show promising suitability to audio recognition tasks and might be an interesting consideration for future work extensions on this topic.

Since the choice of the ML model architecture has a major impact on the resulting outcome, \cite{palanisamy2020rethinking} have compared different pre-trained models for training. For example, Inception\cite{szegedy2015going}, ResNet\cite{he2016deep}, and DenseNet \cite{huang2017densely}. In each case, five independently trained models were merged for predictions using an Ensemble. 
In doing so, they were able to demonstrate that CNN models can predict bounding values of energy distribution in spectrograms (in general) to classify them. Ensembling made it possible to increase the performance of the predictions.\cite{dietterich2000ensemble}, for instance, also uses a ensemble voting system to determine the final predictions of the different models. \cite{nanni2021ensemble} use such an ensemble approach for the classification of audio data with a focus on surveying different, popular CNN architectures.
Kong et al.\cite{kong2020sound} describe sound event detection on weakly labeled data using CNN and CRNN with automatic boundary optimization. Here, an audio signal is also divided into segments and the CNN model is trained on them. This determines whether certain audio signals, based on their segments, contain usable information or not. The segments are given all the properties of the original data, such as the label. Subsequent classification checks whether the original label matches the prediction. MEL spectrograms were used as input to the CNN. We have modeled our binary classifier on sample segments with a similar intent.


\section{\uppercase{Primate Dataset (ComparE)}}
\label{sec:primate_dataset}

\begin{figure}
    \centering
    \subfloat[][Distribution of the audio data]{\label{fig:class_distribution}
    \includegraphics[trim={0 0 1.5cm 1.5cm},clip,width=0.5\linewidth]{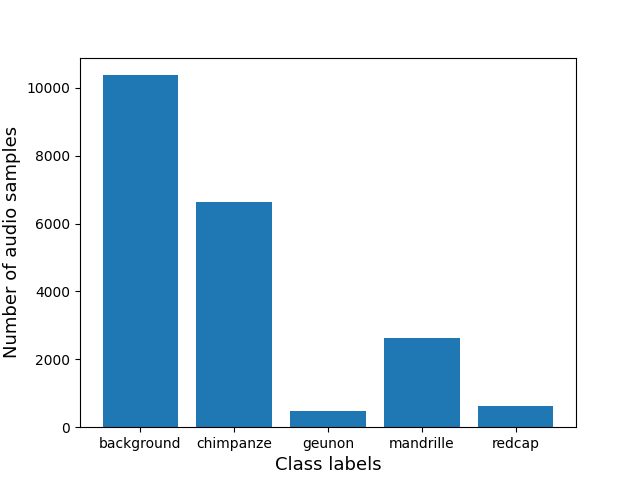}}
    \subfloat[][Size of recordings in seconds]{\label{fig:duration_distribution}
    \includegraphics[width=0.5\linewidth]{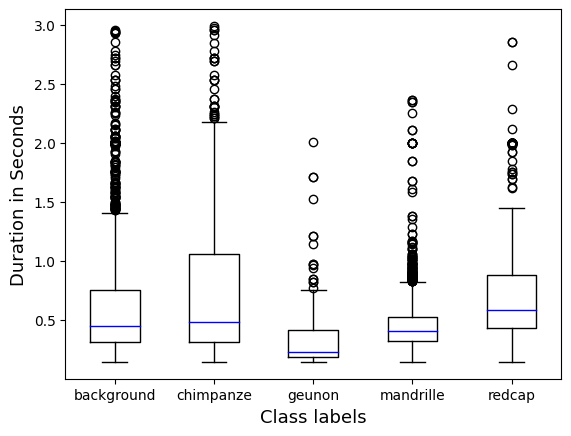}}
    \caption{In \ref{fig:class_distribution} the distribution of the audio data of the individual classes in the entire data set is depicted. In total there are 20,000 data samples. In \ref{fig:duration_distribution} the size of recordings in seconds for audio samples of the respective classes in the entire data set is shown. The duration of the individual data is between 0.145 and 3 seconds.}
\end{figure}

The `Central African Primate Vocalization dataset'~\cite{zwerts2021introducing} was collected in a primate sanctuary in Cameroon. The samples contain recordings of the five classes; Chimpanzees, mandrills, collared mangabeys (Redcapped Mangabeys), guenons, as well as forest-noise (`background' or `BKGD'). 
The rate of primates samples to background sound samples is 1:1 (10,000 each, in total 20,000 samples). 
As shown by Figure~\ref{fig:class_distribution}, unfortunately the data classes are highly imbalanced, which makes the classification more difficult.

Another problem lies within the distribution of the audio durations per sample (cf. Figure~\ref{fig:duration_distribution}); Data points appear with lengths between 0.145(!) and 3 seconds, with samples from guenons and mandrills having particularly short audio durations.
Considering both the imbalanced distribution of the classes and the highly varying audio durations, in addition to the high amount of noisy samples, we have to consider this dataset of advanced difficulty. Correctly classifying the lesser represented classes (guenons and mandrills), will be a major challenge, even for deep learning models.

In addition -- as described by \cite{zwerts2021introducing} -- the so-called \textit{signal-to-noise} ratio for two of the classes (mandrills and collar mangabees) is a particularly significant problem, as both classes are more difficult to distinguish those from general noise in the background. Signal-to-noise ratio (often abbreviated as \textit{SNR} or \textit{S/N}) is a measure that indicates the ratio of a signal to background noise defined by

\begin{equation}\label{formula:SNR}
    SNR_{\text{dB}} = 10\;log_{10}\left(\frac{\text{signal}}{\text{noise}}\right)
\end{equation}

For our purpose, this ratio of energy from the signal in contrast to the noise energy is calculated and then converted to decibels (dB) \cite{hemalatha2015wavelet}.
A low SRN value therefore indicates a high amount of noise in the audio signal relative to the informative signal share and vice versa. Consequently, a high value is desirable most of the time.

Noise in this concept is often regarded as a some kind of static introduced by, e.g., the recording equipment. 
Unfortunately, this is not the kind of noise this dataset is influenced by, rather than natural background noise such as `forest noise', human talking etc.
Background noise, as such, is often mixed with the pure informative signal itself, but also present in samples, that consist of many in-consecutive informative parts (such as complex multipart primate screams).
In addition to the regular classification task, a valid automated approach would have to distinguish between background noise and the actual signal first.
This fact, coupled with the class and length distribution issues discussed above, renders the automated detection and labeling of the ComparE data samples a considerably tricky problem.
\section{\uppercase{Method}}\label{sec:methods}

\begin{figure}
    \centering
    \includegraphics[width=0.5\linewidth]{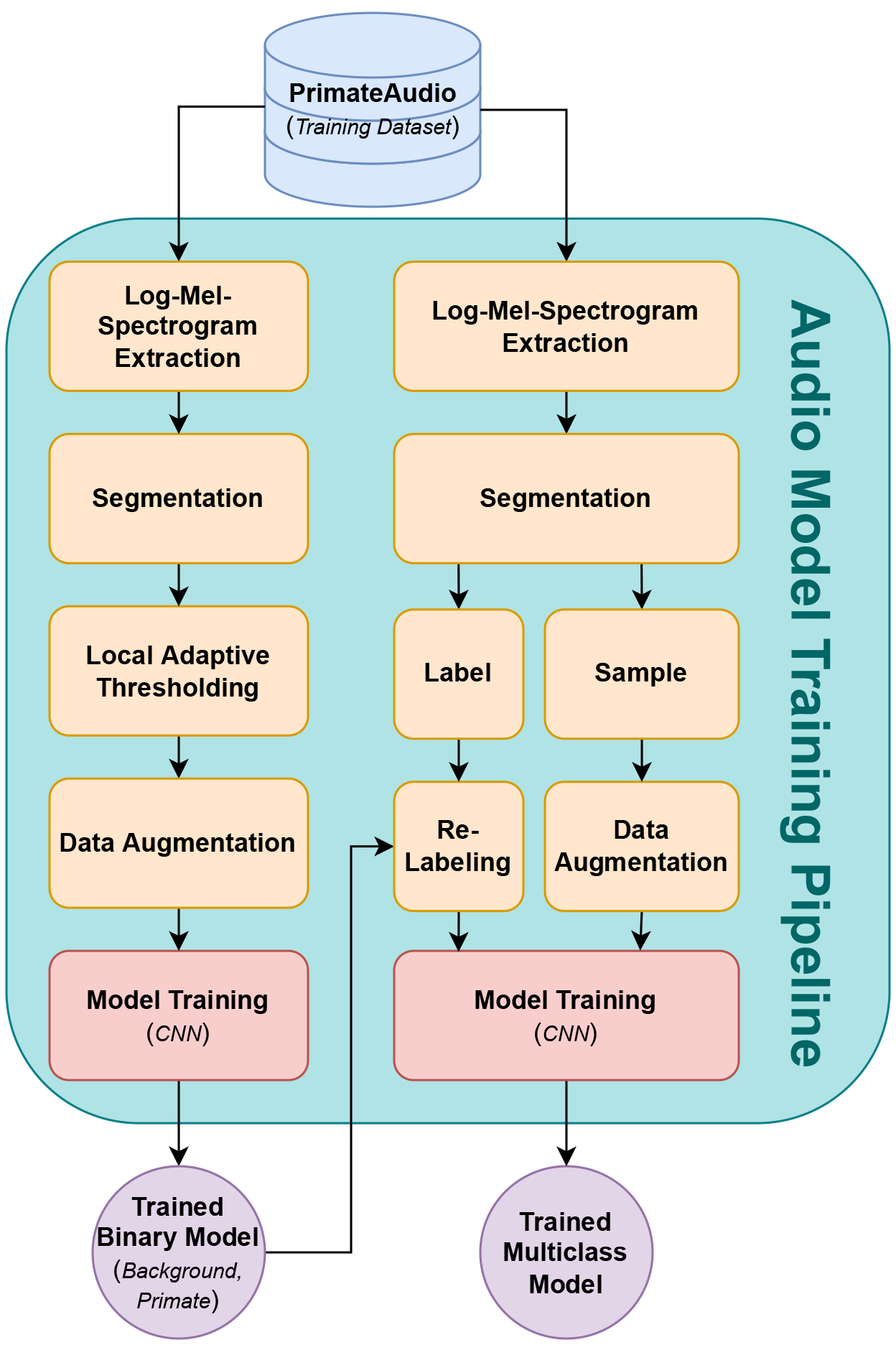}
    \caption{Pipeline for training a multi-class classifier with binary pre-sorting by a binary classifier trained with thresholding. On the left, the pipeline for training a binary model is shown. Here, pre-sorting is used to adjust the label of a segment in the process of multi-class training.}
    \label{fig:pipeline}
\end{figure}

As is, this data set represents a collection of quite undesirable, challenging properties. To address these issues, we utilize a selection of preprocessing methods, like segmentation, thresholding, data augmentation and, most importantly, a pre-sorting approach on the equal length segments of every data sample. 
As is common practice with audio-related ML classification, we first convert the recordings into MEL spectrogram images, by converting the signal to the frequency domain using a (Fast) Fourier Transform (FFT) \cite{bracewell1986fourier}. Spectrograms are particularly useful for audio analysis because they contain a lot of information about the audio file visually and can be used directly as input for our convolutional neural network (CNN) classifier. The formula for converting a FFT spectrogram to the MEL representation is given by:
\begin{equation}\label{formula:MEL}
    MEL(f)=2595\;log_{10}\left(1+\frac{f}{700}\right)
\end{equation}
where \textit{f} is the spectrogram frequency in Hertz.
Using image data instead of pure audio also offers the advantage of many easy and fast to use data augmentation methods, fast training, easy low-level feature (LLF) extraction (through the MEL spectrogram) and small model sizes. A schematic overview of our complete training pipeline is shown in Figure~\ref{fig:pipeline}.

\subsection{Segmentation}\label{subsec:meth_segmentation}
\vspace{-6pt}
Since the lengths of the individual audio data differs greatly, we segment the data samples. This not only increases our net-amount of data, but also ensures that training data passed to the model is of uniform length. The samples are divided into sections of equal lengths, directly on the MEL spectrograms image of the audio data. (Segment length is a variable hyperparameter.) The individual segments inherit the label of the original audio file. Certain windows can thereby contain no primate sounds but are labeled otherwise and vice versa. We remedy this problem afterwards by pre-sorting, which is described in more detail in Section~\ref{subsec:meth_pre-sorting}. Note that the predetermined segment length for binary training must be identical to that of multi-class training, otherwise the size of the network for predictions will not match that of multi-class training. In this process, the augmentation of the data is performed on the individual segments of the audio data.

\subsection{Thresholding}\label{subsec:meth_thresholding}
\begin{figure}
     \centering
     \subfloat[][Windows]{\includegraphics[trim={5pt 0 50pt 0} ,clip,width=.5\linewidth]{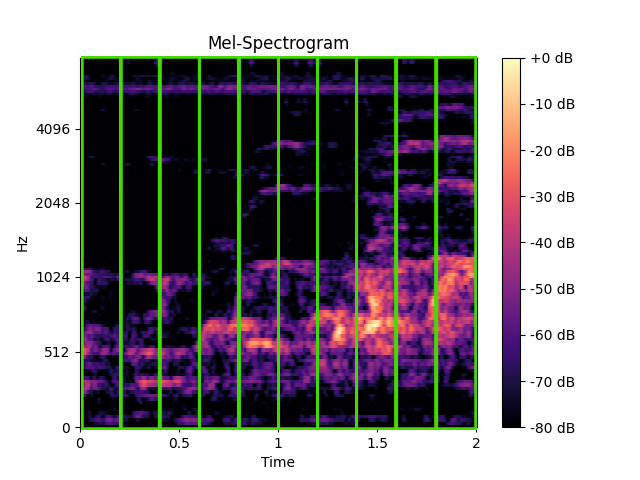}\label{fig:thresholding_windows}}
     \subfloat[][Thresholding applied]{\includegraphics[trim={5pt 0 50pt 0},clip,width=.5\linewidth]{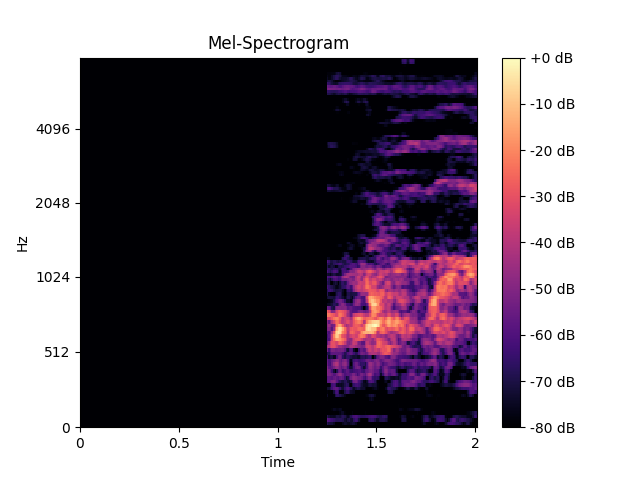}\label{fig:thresholding_applied}}
     \caption{MEL spectrogram thresholding demonstration. We iterate over the samples with a separate thresholding window of a predefined size and sum the MEL frequency of the respective window. The signal frequency is represented by color. The brighter the color in the MEL spectrogram, the stronger is also the expression of the intensity of the signal in this frequency range. The color distribution is indexed in decibels (cf. color scale on the right).}
     \label{fig:1-2}
\end{figure}

\vspace{-6pt}
Thresholding is a technique where we omit sections of a sample, for which the MEL signal falls below an assigned threshold, to ensure that learning can be done on data that contains the most relevant information possible.
Since the audio for the dataset was recorded in the wild, there are many noises of varying intensity present in the recordings. Because of the high SNR variance, even between the primate classes, it would not make sense to use a global threshold. 
To counteract this, \textit{local adaptive thresholding} \cite{thresh} is applied to all sample segments by iterating over the segment with a separate thresholding window (size determined as hyperparameter) and summing the MEL frequency of the respective window. This adds noise and other interfering sounds to the relevant sounds, so that irrelevant sounds are set relative to the actual information content. Subsequently, the threshold (also a hyperparameter) is set relative to the normalized sum of the individual windows of the MEL spectrogram, e.g, the same threshold value is lower for more quiet samples. Segment windows below a calculated threshold are blackened out accordingly so that background noise is removed as far as possible (e.g., Figure~\ref{fig:1-2}). 
The threshold is only intended to differentiate between background noise or a primate cry in a segment. Since this is not relevant for the primary task of multi-class classification, we only use the thresholding for the binary training.


\subsection{Data Augmentation}\label{subsec:meth_data_aug}
\vspace{-6pt}
If a model is trained with an unbalanced data set, it will also be biased in its predictions so that the evaluation classification accuracy differs significantly from the training accuracy. This pitfall is commonly known as overfitting.
To further counteract the class distribution imbalance already shown in Figure~\ref{fig:class_distribution}, there are two common approaches; \textit{Under sampling} instances of the most frequently occurring class – not ideal considering we already do not have that many samples to train on. Nonetheless, to ensure a balanced training process, we use a \textit{Weighted Random Sampler} that draws samples proportional to their class size. 
\textit{Oversampling}, on the other hand, generates synthetic data of underrepresented classes to match the number of the most frequently occurring class. Thus, several instances of data augmentation are undertaken in this work, in addition to Weighted Random Sampling. In general, data augmentation takes advantage of the assumption that more information can be extracted from the original data to enlarge a data set. For \textit{data warping} methods, existing data is transformed to inflate the size of a data set. On the other hand, \textit{methods of oversampling} synthesize entirely new data to add to a data set \cite{shor19}.\\
In this work, augmentations such as noise injection, shifting, pitching, and frequency and time masking are used. Their effectiveness in providing solid performance on this very dataset has been shown in \cite{illium2020surgical,illium2021visual}.
A simple \textit{Loudness} adjustment minimally amplifies the signal across all samples. 
By adding random values (Gaussian / random-normal) to an audio signal (\textit{Noise Injection}) stochasticity is introduced without altering the overall character of the sample. This is done on the background noise samples as well as on the regions of the signal relevant for training. 
\textit{Shifting} shifts the signal left or right by a certain value on the time axis. \textit{Pitching} increases the overall mean of the signal, increasing or decreasing the frequency of the sound. 
\textit{Frequency masking} sets random frequencies on a MEL spectrogram to 0.
The same applies to \textit{time masking}, where time segments that are close to each other on the time axis in the MEL spectrogram have their signal are reduced to 0. Afterwards, all segments are brought to the same length (also called \textit{padding}). Data augmentation is performed on every individual sample-segment of the audio data.


\subsection{pre-sorting With Binary Prediction}\label{subsec:meth_pre-sorting}
\vspace{-6pt}
With all data samples split into multiple segments, some with reduced signal from the thresholding, padding or other modification from the data augmentation, we have gained considerably more data to work with. However, the problem remains that the split has introduced a significant amount of false positive samples, i.e., purely noisy segments that have inherited a primate class label of the whole data point. To correctly re-label all samples, we first pretrain the multi-class predictor for a binary classification task, where we reduce
the prediction of the five given classes to a binary problem (`Background' and `Primates'). We then use this pretrained model to re-label all the (possibly false positive) samples, i.e., actual noise with an old primate label, to the noise label. After all segments are re-labeled, we can change the model's prediction head and continue on with the multi-class training. To clarify, pre-sorting only takes place when the model confidently predicts `Background' for input with a primate label. By pre-sorting the audio segments, the multi-class model can later train more confidently on augmented primate sounds and, crucially, gets pretrained on the general difference of noise versus primate sounds.

\section{\uppercase{Model Architecture}}\label{sec:training}
Following \cite{zwerts2021introducing}, for the classification experiments, we partition the dataset
into training, validation, and test sets with a ratio of 3:1:1.

A CNN was used for both the binary and multi-class classification tasks, the same one, in fact. The Adam Optimizer was chosen as the optimizer, with a learning scheduler with a step size of $100$ epochs and a reduction of the learning rate by $0.05$. 
To calculate the loss, the binary cross entropy-loss (BCE-Loss) was used to train the binary classifier.
Furthermore, we used focal-loss with a value of $2$ for $\gamma$ for the multi-class approach, a dynamically scaled cross entropy loss which is suited for datasets with high-class imbalance. Batch normalization, as well as dropout, were not used for training the binary model.
For (pre-)training, various hyperparameter configurations were examined and compared using a mix of Unweighted Average Recall (UAR), as well as Accuracy and F1-Score.
The F1 score \cite{muc4} was used to determine the most informative model to be used as a binary classifier. 
As mentioned above, for the pretrained generalization to be transferable to the multi-class task, it is crucial that the model is trained with the identical segment length as the binary model used for training and pre-sorting. 
A collection of the (final) parameters we used for training is shown in Table~\ref{table:params}.

\begin{table}[hpbt]
    \centering
    \begin{tabular}{ l|r|c|l|r} 
     \textbf{Parameter} & \textbf{Val.} && \textbf{Parameter} & \textbf{Val.} \\ \hline\hline
     Epochs & 200                       && Batch size & 32 \\ 
     Learning Rate & 1e-4               && Sample Rate & 16 K \\
     \textit{n} FFT's & 1024            && Hop size & 128 \\
     \textit{n} MEL Bands & 128         && Segment length & 0.7 s \\
     Threshold & 0.3                    && Gap Size & 0.4 s\\ 
     Randomness & 0.5                   && Loudness & 0.3 \\
     Shift & 0.3                        && Noise & 0.3 \\
     Mask & 5                           && Dropout & 0.2 \\
    \end{tabular}
    \caption{Hyperparameters and their set values for binary, as well as multi-class training. 
    The sample rate is given in Hertz and the segment length, as well as the window size of the threshold (Threshold Gap Size) in seconds. 
    Randomness indicates the probability with which the respective augmentations, such as loudness, shift, noise and masking, are applied. A value of $50\%$ is used for each of these augmentations. Consequently, it is also possible that, for example, no loudness is used, but shift and noise are. 
    The hyperparameter for the time or frequency masking is specified here as mask. 
    This value determines how many adjacent samples on each of the two axes should be converted to zero. 
    Note that the binary model was only trained for $150$ epochs.}
    \label{table:params}
\end{table}


The CNN model used is composed of five hidden layers. Each of the hidden layers consists of a convolutional layer, batch normalization, the activation function ReLU, max-pooling and dropout. The convolutional layer has a window or kernel size of $3x3$, stride of size $1$ and padding with the value $2$. The following batch normalization corresponds to the size of the output of the previous convolutional layer. Thereupon, a max-pooling layer with a window size of $2x2$ is applied, with a subsequent dropout. The max-pooling makes it possible for the model to look more closely at contrasts in the MEL spectrogram and thus better detect relevant information in audio data. Finally, in the final dense layer, all values from the last hidden layer are taken and reduced to one dimension so that they can be passed to a suitable activation function. For the multi-class model, a typical Softmax activation function was used for multi-class classification, which maps the output to a probability distribution of the individual labels. For the binary classifier, a sigmoid activation function was applied instead.

\section{\uppercase{Experiments}}
\label{sec:experiments}
Our experiments involve the following models: A multi-class model, to which neither thresholding nor pre-sorting was applied as a baseline (only audio data into MEL spec. conversion, segmentation \& data augmentation); two binary models (w/wo thresholding); and two multi-class models involving the binary models for pre-sorting.
Finally, we investigate the impact of the hyperparameters segment length, hop length, as well as the number of FFT's.

\subsection{Evaluation}
\vspace{-6pt}
The results of each training run are measured using the expected and actual prediction of a model. Based on these results, the metrics Unweighted Average Recall (UAR), as well as Accuracy and F1-Score are calculated. Accuracy and UAR are defined as 
\begin{equation}
    \text{Accuracy} = \frac{TP + TN}{TP + TN + FP + FN}
\end{equation}
and
\begin{equation}
    \text{UAR} = \frac{1}{N}\displaystyle\sum_{n=1}^{N} \frac{TP_n}{TP_n+FN_n}
\end{equation}
where TP, TN, FP, FN, P, N are \textit{True Positives}, \textit{True Negatives}, \textit{False Positives} and \textit{False Negatives}, \textit{Positives (Total)}, \textit{Negatives (Total)}, and number of classes ($n$) respectively.
These metrics indicate how precisely a model has been trained under certain conditions. For binary training, additional mismatch matrices are used to be able to determine exactly which classes have been incorrectly labeled. Thus, a binary model with an insufficient mismatch matrix could be discarded. Similarly, confusion matrices are used to observe exactly which classes have been incorrectly assigned in multi-class training. 
To serve as comparison, the baseline model was trained with $200$ epochs and achieved an Accuracy of $95.23\%$ after $125$ epochs, and a UAR of $82.86\%$ on the test set. Guenons were incorrectly labeled as background in $18.9\%$, chimpanzees in $17.99\%$, and collar mangabees in $19.05\%$ of cases.

\begin{figure}
    \centering
    \subfloat[][Confusion matrix of the baseline]{\label{fig:cm_baseline}
    \includegraphics[width=0.5\linewidth]{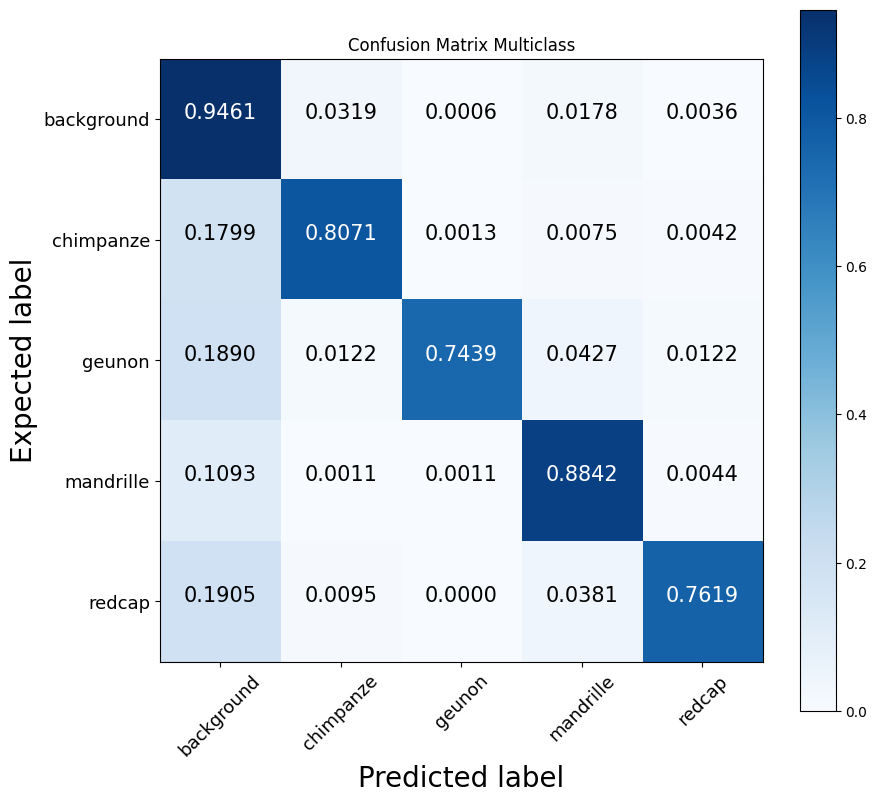}}
    \subfloat[][Mismatch Confusion Matrix]{\label{fig:cm_missmath_no_thresholding}
    \includegraphics[width=0.5\linewidth]{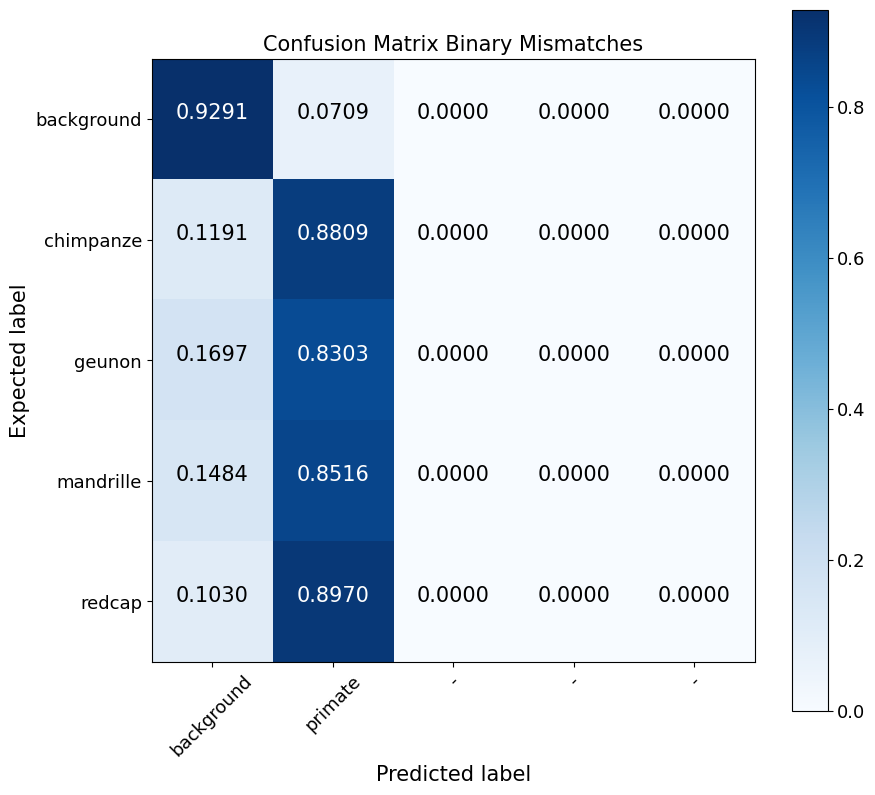}}
    \caption{Confusion matrix of the baseline used for this work (\ref{fig:cm_baseline}). The X-axis describes the predicted labels of the classifier, whereas the Y-axis describes the actual true labels. The diagonal represents the correctly predicted labels of the individual classes. Mismatch Confusion Matrix (\ref{fig:cm_missmath_no_thresholding}) for the trained binary model without thresholding. The second column indicates which primates were correctly classified. were correctly classified. The first column represents which audio data was detected as noise or background.}
\end{figure}

\subsection{Binary pre-sorting}
\vspace{-6pt}
We trained two binary models, one with and one without thresholding. The goal was to compare the two approaches under the same initial conditions. The binary model without thresholding achieved an F1 score of $90.2\%$ on the validation dataset. Other metrics scored between $90\%$ and $91\%$. Figure~\ref{fig:cm_missmath_no_thresholding} shows the resulting mismatch confusion matrix, which clearly indicates which classes have been mislabeled, which classes have been mislabeled. Thereby, guenons with $16.97\%$ and mandrills with $14.84\%$ were incorrectly labeled as background. 

A binary model was then trained with thresholding. Different window sizes for thresholding were investigated to counteract the SNR problem. For this, two binary models were trained with a threshold window size of $0.2$ seconds and $0.4$ seconds, and a threshold of $30\%$ for each. The remaining hyperparameters for the training can be taken from the Table \ref{table:params}. 


The binary model with thresholding and a window size of $0.2$ seconds achieved a UAR of $85.94\%$, while the second model with a size of $0.4$ seconds was able to produce a much better UAR of $89.15\%$. The miss-classifications with a small window size are very high for chimpanzees with $21.49\%$, guenons with $30.3\%$ and collar mangabees of $38.21\%$. The mismatch matrix for thresholding with a window size of $0.4$ seconds is significantly lower in values than the matrix without thresholding (cf. Figure~\ref{fig:cm_missmath_no_thresholding}) 
Consequently, the model with the window size of $0.4$ seconds was used for the pre-sorting of the multi-class model.


\subsection{Multi-Class Classifier}
\begin{figure}
     \centering
     \subfloat[][with thresholding]{
        \includegraphics[width=0.5\linewidth]{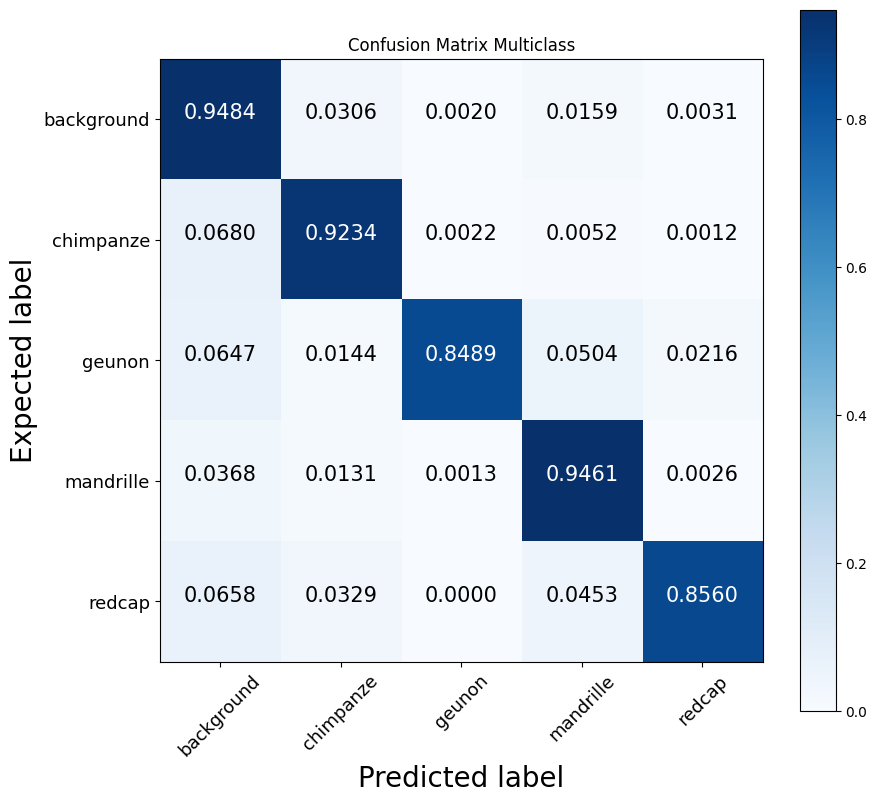}
        \label{fig:mc_confusion_matrix_with_threshold}
    }
     \subfloat[][without thresholding]{
        \includegraphics[width=0.5\linewidth]{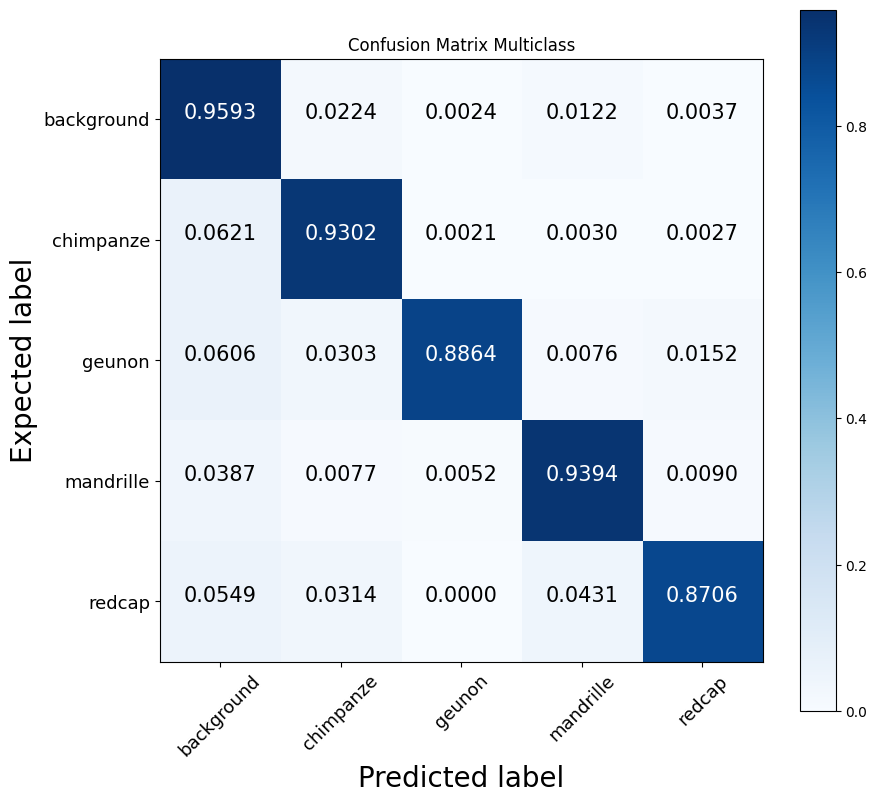}
        \label{fig:mc_confusion_matrix_without_threshold}
    }
     \caption{(Re-labeled) Multi-class confusion matrices pretrained with and without thresholding}
     \label{fig:mc_confusion_matrix}
\end{figure}

\vspace{-6pt}
Finally, two multi-class models were trained with binary pre-sorting. For this step, one binary model with thresholding and one without were examined (cf.~Figure~\ref{fig:mc_confusion_matrix}). Both models, in fact, produce quite competitive results, both in terms of accuracy and UAR. We observe, however, that the model without thresholding performs notably better than the one with this technique. The high SNR in many of the class samples is the likely culprit, as thresholding (even the locally adaptive variant), has difficulties setting clean threshold values. We have collected an overview of the test scores of works attempting the ComparE dataset challenge (including our models) in Table~\ref{table:mc_results}.


\begin{table*}[ht]
\centering
\begin{tabular}{ l|c|c|c|c } 
    \textbf{multi-class-Model} & \textbf{Acc. Val} & \textbf{Acc. Test} & \textbf{UAR Val} & \textbf{UAR Test} \\ \hline\hline
    Baseline (our) & 95.69 & 95.23 & 85.48 & 82.86 \\
    Re-Label (our) w/o Threshold & \textbf{97.83} & \textbf{97.81} & \textbf{92.84} & \textbf{91.72} \\ 
    Re-Label w/ Threshold (our) & 97.54 & 97.48 & 92.29 & 90.46 \\ \hline
    ComParE~21 Baseline (Fusion)\textsuperscript{1} & - & - & - & 87.5 \\
    CNN10-16k\textsuperscript{2}~\cite{pellegrini21_interspeech} &-&-& \underline{93.6} & \underline{92.5}\\
    ComParE + FV (FBANK) + auDeep\textsuperscript{3} &-&-& 88.2 & 89.8\\
    ComParE + FV (FBANK)\textsuperscript{3} &-&-& 87.5 & 88.8 \\
    DeepBiLSTM\textbf{\textsuperscript{4}} &-&- & 89.4 & 89.3  \\
    ResNet22-16k\textsuperscript{2} & - & - & 92.6 & - \\
    Vision Transformer\textsuperscript{5} & - & - & 91.5 & 88.3 \\
    \hline
\end{tabular}
\caption{Results of the different multi-class models on the validation and test data set. The baseline, a multi-class model with pre-sorting but without thresholding, and a multi-class model with pre-sorting and thresholding were compared. Thresholding was used only for training the binary relabeler. \\
\textbf{Note:} Additional scores are cited as reported by the authors: \\
    \textbf{\textsuperscript{1}}Schuller et al., 2021, \\
    \textbf{\textsuperscript{2}}Pellegrini, 2021, \\
    \textbf{\textsuperscript{3}}Egas-López et al., 2021, \\
    \textbf{\textsuperscript{4}}Müller et al., 2021, \\
    \textbf{\textsuperscript{5}}Illium et al., 2021.}
\footnotetext{\cite{schuller2021interspeech,pellegrini21_interspeech,egaslopez21_interspeech,muller2021deep}}
\label{table:mc_results}
\end{table*}

\section{\uppercase{Conclusion}} \label{sec:conclusion}
Let's summarize: In this work, we have presented and evaluated a dataset pre-sorting approach as a pre-processing step for training of CNN-based classifiers. We have tested our training pipeline – including segmenting, local adaptive thresholding, data-augmentation and pre-sorting – on the challenging ComparE dataset of primate vocalizations and observed competitive results to other participants of the challenge. We managed to show that the binary-pre-sorting approach significantly outperformed the baseline without this step, indicating the value of splitting and pre-sorting difficult, i.e., high-variance, imbalanced data, resulting in a more confident final multi-class classification. Compared to our basic baseline, the UAR could be increased from $82.86\%$ to $91.72\%$. As a bonus, this binary differentiation task between background noise and any primate sound, serves a beneficial pre-training for the actual prediction task, as the model can start the training on a general notion `primate or noise' and then fine-tune its weights on the multiple-primate task.

We have also observed that thresholding, as a pre-processing step for training of the binary pre-sorting model, has had no significant positive effect on the resulting test accuracy metrics (on the contrary). For the final model (with binary-pre-sorting) we observed an UAR of $89.15\%$, which is better than the baseline but worse than the model without thresholding. The reason for this may lie in the data itself, the weakly labeled samples, or the highly imbalanced (class and length) distribution. The low duration of some audio samples, in particular, additionally prevents the thresholding from finding beneficial application. It is important to mention that especially the low SNR of mandrills and collar mangabees in the audio data set are responsible for the fact that the affected classes are difficult to distinguish from background noise. While thresholding may certainly offer promising application on other data sets, for high SNR ratio data like the ComparE dataset, we find the suitability rather pool.

For future work, the effects of certain hyperparameters and their dependencies could be investigated in more detail, e.g., like we have shown that the number of FFT's to hop length significantly affected the performance of binary training.
Additionally, it is unclear how promising the binary presort works on other datasets. Therefore, it would be possible to investigate and compare this approach on other data sets. to compare. Furthermore, it would be conceivable to train an additional model on the noisy segments. on the noisy segments. This would make it possible to use another model for pre-sorting to provide more meaningful data and labels to the multi-class model. Especially for current emerging technologies like quantum machine learning, binary presorting could help to decrease the amount of input data that is needed for classification tasks. Finally, it might also make for interesting future work to examine the transfer-ability of pretrained models on normal images to spectrogram imagery and related audio tasks.

\section*{\uppercase{Acknowledgements}}
This paper was partially funded by the German Federal Ministry of Education and Research through the funding program ``quantum technologies --- from basic research to market'' (contract number: 13N16196).

\bibliographystyle{splncs04}
\bibliography{main}
\end{document}